# Frequency-tunable circular polarization beam splitter using a graphene-dielectric sub-wavelength film

Tuo Chen[1,2] and Sailing He [1,3 *]

[1]State Key Laboratory of Modern Optical Instrumentations, Centre for Optical and Electromagnetic Research, JORCEP, Zhejiang University, Hangzhou 310058, China

[2]Department of Physics, Zhejiang University, Hangzhou 310058, China

[3]Department of Electromagnetic Engineering, School of Electrical Engineering, Royal Institute of Technology, S-100 44 Stockholm, Sweden

[*] Corresponding author's email: sailing@kth.se

**Abstract**: Manipulating the circular polarization of light is of great importance in chemistry and biology, as chiral molecules exhibit different physiological properties when exposed to different circularly polarized waves. Here we suggest a graphene/dielectric-stacked structure, which has both the properties of a epsilon-near-zero material and the high Hall conductivity of graphene. The proposed sub-wavelength structure demonstrates efficient manipulation of circular polarization properties of light. In a quite broad frequency range and at a large oblique incidence angle, the present magnetically active structure is transparent for one circularly polarized wave, and opaque for another. Such an effect can be further tuned by changing the magnitude of the applied magnetic field and chemical potential of graphene.

The capability to manipulate the circular polarization of light is one of essential optical applications of molecular biology, medical science and analytical chemistry. Most biomolecules, including the basic block of life, DNA, are chiral, which are sensitive to optical stimuli and thus behave differently when exposed to left-handed circular (LCP) waves and right-handed circular (RCP) waves. The most commonly used method for manipulate circular polarization of light is to use half- or quarter-wave plates with a retardation effect in an anisotropic crystal. However, the small differences in permittivity between the corresponding crystallographic directions require a thick retardation distance, resulting in dozens of wavelength thicknesses. On the other hand, progress with chiral heterostructrues [1-3] provides an unprecedented opportunity to manipulate the circular polarization of light by tailoring the refractive index of the material. Nevertheless, most chiral metastructures are based on Bragg reflection and localized plasmons, which lead to a narrow frequency band. Although the spectrum can be broadened by stacking multiple polarizers or introducing a gradient helical pitch [4], the relative strength of different circularly polarized waves and the working spectrum can not be tuned without changing the physical structure of the chiral material.

Graphene nanophotonics has attracted considerable interest due to its unique optical properties [5,6] and many related applications [7-9]. Apart from being the thinnest exciting material, graphene is attractive for its physical flexibility, high electron mobility and the possibility of controlling its carrier concentration via external gate voltages or chemical doping. Another promising application of graphene is in magneto-optics. The ability to create a giant Faraday rotation angle has been indentified experimentally in both single-layer and multilayered continuous graphene [10]. The large Hall conductivity [11], which produces extremely significant Faraday rotation originating from the cyclotron effect in the classical region and the inter-laudau-lever transitions in the quantum regime, inevitably affects RCP and LCP waves differently, both in amplitude and phase. Epsilon-near-zero (ENZ) material also exhibits highly unusual and intriguing optical properties. One of these properties is the near-zero phase delays while a wave propagates though such an ENZ hyperbolic material, which has been proposed for extraordinary transmission in the bend structure [12]. ENZ materials were also employed for radiation pattern control [13,14], and perfect absorption [15]. Combined with magneto-optical materials, ENZ materials may provide a mechanism for enhancing the effects of nonreciprocity and time-reversal symmetry breaking [16,17].

In this paper, by employing the properties of ENZ materials and the high Hall conductivity of graphene, we propose a fast tunable magneto-optical sub-wavelength structure that demonstrates efficient manipulation of circular polarization properties of light. The magnetically active structure is transparent for one circularly polarized wave, and opaque for another. In a fairly broad frequency range and at a large oblique incidence angle, such a polarizing effect can be realized and tuned by changing the magnitude of the applied magnetic field or the chemical potential of graphene.

## RESULT AND DISCUSSION

To model one-atom-thick graphene in a macroscopic electromagnetic description, the graphene can be treated as an ultrathin film with thickness $\Delta$ ($\Delta \to 0$). The equivalent parallel complex permittivity of the $\Delta$-thick graphene layer is $\xi_g = i\sigma/(\xi_0 \omega \Delta)$. In the presence of a static magnetic field $B$ applied perpendicular to its surface, graphene is gyrotropic and its anisotropic permittivity can be described by a tensor:

$$\bar{\bar{\xi}}_g = \begin{pmatrix} \xi_{gd} & \xi_{god} & 0 \\ -\xi_{god} & \xi_{gd} & 0 \\ 0 & 0 & \xi_z \end{pmatrix} \quad (1)$$

where $\xi_{gd} = i\sigma_{xx}/(\xi_0 \omega \Delta)$ and $\xi_{god} = i\sigma_{yx}/(\xi_0 \omega \Delta)$. $\sigma_{xx}$ and $\sigma_{yx}$ are the longitudinal conductivity (along the E-direction) and Hall conductivity (along the direction perpendicular to E), respectively.

Consider a graphene layer which is biased by a static magnetic field perpendicular to the surface, and its electronic property can be characterized by the surface conductivity, which can be modeled by the

Kubo formula [16] through a quantum mechanical analysis, as

$$\sigma_{xx}(\omega,B) = \frac{e^2 v_f^2 |eB|\hbar(\omega+i2\Gamma)}{i\pi} \sum_{n=0}^{\infty} \{\frac{1}{M_{n+1}-M_n} \times [f_d(M_n)-f_d(M_{n+1})+f_d(-M_{n+1})-f_d(-M_n)]\}$$
$$\times \frac{1}{(M_{n+1}-M_n)^2 - \hbar^2(\omega+i2\Gamma)^2} + (M_n \to -M_n)\}$$

(2a)

$$\sigma_{yx}(\omega,B) = -\frac{e^2 v_f^2 eB}{\pi} \sum_{n=0}^{\infty} [f_d(M_n)-f_d(M_{n+1})-f_d(-M_{n+1})+f_d(-M_n)]$$
$$\times [\frac{1}{(M_{n+1}-M_n)^2 - \hbar^2(\omega+i2\Gamma)^2} + (M_n \to -M_n)]$$

(2b)

where $v_f \approx 10^6$ m/s is the Fermi velocity, $M_n = \sqrt{n}L$ are the Laudau energy levels, and $L^2 = 2\hbar eB v_f^2$ is the Laudau energy scale. $f_d(\xi) = [exp(\xi-\mu_c)/(k_B T)+1]^{-1}$ is the Fermi-Dirac distribution, $\mu_c$ is the chemical potential, $k_B$ is the Boltzmann constant, $T$ is the environment temperature, $\hbar = h/(2\pi)$, $h$ is the Planck constant, and $\Gamma$ is the phenomenological scattering rate.

The total conductivity depends on two parts of electron transition, namely, the intraband transition and interband transition. The former transition involves energy levers in the same band (both in the conduction band or the valence band), and the latter transition involve levels in different bands. As a result, interband transitions occur essentially at frequencies $\hbar\omega \geq 2\mu_c$. The spectral range we study in this paper is consistent with the restriction $\hbar\omega < 2\mu_c$. Therefore the results of this paper are derived by including the intraband term in (1). Combining this with the fact that the transitions between the levels around $\mu$ are the strongest ones, it shows that the conductivities follow the Drude model form [17]

$$\sigma_{xx} = \sigma_0 \frac{1-i\omega\tau}{(\omega_c\tau)^2 + (1-i\omega\tau)^2} \quad (3a)$$

and

$$\sigma_{yx} = \sigma_0 \frac{\omega_c\tau}{(\omega_c\tau)^2 + (1-i\omega\tau)^2} \quad (3b)$$

where

$$\sigma_0 = \frac{2e^2\tau}{\pi\hbar^2} k_B T \ln(2\cosh\frac{\mu_c}{2k_B T}) \quad (3)$$

In this semiclassical expression for the conductivity, $\tau=1/(2\Gamma)$ is the scattering time and $\omega_c$ is the cyclotron frequency, which corresponds to the difference between the neighbor Laudau energy around the Fermi lever.

$$\omega_c = \frac{M_{n+1} - M_n}{\hbar} \approx \frac{L^2}{2\hbar\mu_c} = \frac{eBv_f^2}{\mu_c} \quad (4)$$

Since a DC mobility of high-quality suspended graphene is expected to be 200000 $cm^2V^{-1}s^{-1}$ [18,19], and $\mu > 60000$ $cm^2V^{-1}s^{-1}$ (corresponding to the scattering time of about 1.8 ps) in graphene on a hexagonal boron nitride (h-BN) substrate was experimental achieved [20], it is possible to safely assume $\tau=2$ ps throughout the study in our paper.

The schematic diagram of the suggested structure is shown in Fig. 1, which consists of periodically alternating graphene and dielectric layers.

The dielectric layer has thickness $t_d=300\ nm$ and permittivity $\xi_d=3$. A static magnetic field $B$ is applied perpendicularly to the interface of such two materials. The graphene/dielectric stack can be described as a hyperbolic metamaterial with effective constitutive parameters, since the thickness of the dielectric layer $t_d$ is much smaller than the wavelength. There are three effective permittivity parameters, the parallel diagonal and off-diagonal elements $\xi_p$ and $g$ for the electric fields along the graphene/dielectric interface, and $\xi_t$ for the electric field perpendicular to the surface.

$$\bar{\bar{\xi}}_{ave} = \begin{pmatrix} \xi_p & -ig & 0 \\ ig & \xi_p & 0 \\ 0 & 0 & \xi_t \end{pmatrix} \quad (6)$$

According to the effective medium theory, we have $\xi_p = f\cdot\xi_{gd}+(1-f)\cdot\xi_d$, $g= f\cdot i\cdot\xi_{god}$ and $1/\xi_t = f/\xi_z+(1-f)/\xi_d$, where $f=\Delta/(\Delta+t_d)$ is the filling factor of graphene. When $\Delta$ is infinitely small, the three effective permittivity parameters can be reduced to $\xi_p = f\cdot\xi_{gd}+\xi_d$, $g=f\cdot i\cdot\xi_{god}$ and $\xi_t=\xi_d$. Since $\xi_g=i\sigma/(\xi_0\omega\Delta)$ and $f=\Delta/t_d$, we can easily get $\xi_p=\xi_d+i\sigma_{xx}/(\xi_0\omega t_d)$ and $g=-\sigma_{yx}/(\xi_0\omega t_d)$. Given a frequency in the THz range, without properly tuning the magnitude of the applied static magnetic field nBor by properly setting the chemical potential $\mu_c$, which relates to electrostatic biasing and chemical doping, we can achieve near-zero permittivity for $\xi_p$. The dependence of the frequency of ENZ with both magnetic field $B$ and chemical potential $\mu_c$ is rendered in Fig 2a. Throughout the paper, the thickness of the dielectric layer is assumed to be 300 *nm*, which is far below the wavelength corresponding to dozens of THz (the frequency of ENZ is shown in Fig 2a), and the effective medium method used here is valid.

Optical properties of such a graphene/dielectric stack magnetized along the z-axis are characterized by the permittivity matrix (6). Let us consider the incident electromagnetic wave with the form:

$$E = (A\vec{v}_{RH} + B\vec{v}_{LH})\exp(i[\gamma x + \beta z - i\omega]) \tag{7}$$

Where $\vec{v}_{RH}$ and $\vec{v}_{LH}$ correspond to the right-handed and left-handed polarization waves, respectively. $\theta$ is the angle of incidence, and the transverse wave vector $\gamma = k_0 \sin(\theta)$.

In the magnetized graphene/dielectric layer:

$$E = (\kappa_1 \vec{v}_1 \exp(i\beta_+ z) + \kappa_2 \vec{v}_2 \exp(-i\beta_+ z) + \kappa_3 \vec{v}_3 \exp(i\beta_- z) + \kappa_4 \vec{v}_4 \exp(-i\beta_- z))\exp(i[\gamma x - i\omega]) \tag{8}$$

Note that the electromagnetic field in the magnetic layer consists of four plane waves with different polarization states, which could contain both LH and RH polarization components, especially in oblique incidence. With the Fresnel equation, $\beta_\pm$ and $\vec{v}_i$ in the magnetic media can be calculated as:

$$\beta_\pm^2 = \frac{\gamma^2(\xi_t - \xi_p) \pm \sqrt{4\xi_t g^2 k_0^2(\xi_t k_0^2 - \gamma^2) + \gamma^4(\xi_p - \xi_t)}}{2\xi_t} + \xi_p k_0^2 - \gamma^2 \tag{9a}$$

$$\vec{v}_i = \begin{pmatrix} 1 \\ \dfrac{igk_0^2}{\beta_i^2 + \gamma^2 - \xi_p k_0^2} \\ \dfrac{\beta_i \gamma}{\gamma^2 - \xi_t k_0^2} \end{pmatrix} \quad (9b)$$

For the plane wave propagating along the magnetization direction, we have $\gamma = 0$, and in the forward direction equation (9a) reduces to a well known relation for the corresponding wavevectors $\beta_\pm = k_0\sqrt{\varepsilon_p \pm g}$. The plus sign is for a forward propagating RCP wave, and the minus sign is for a forward propagating LCP wave. As the convention for the polarization handedness, the backward propagating waves have the wave vectors, $\beta_\pm = -k_0\sqrt{\varepsilon_p \pm g}$. The plus sign is for a LCP wave, and the minus sign is for a RCP wave. The eigenmodes in such a multi-layered structure are two circularly polarized wave $v_\pm = (1, \pm i, 0)$, corresponding to wavevector $\beta_\pm$. Assuming a lossless condition, the value of $\varepsilon_p$ goes to zero, and the wave vectors $\pm\beta_+$ are real. Therefore the medium is transparent to forward RCP and backward LCP. The other wave vectors $\pm\beta_-$ are purely imaginary. This means that the medium is opaque for the forward LCP and backward RCP.

In the graphene/dielectric alternately stacked material, one can achieve a real part of the diagonal element $\xi_p$ in matrix (6) near zero, while its imaginary part remains quite small. As magnetically biased graphene possessing giant gyrotropic properties at THz frequencies, we can also achieve a large off-diagonal element *g* compared to the convention material. In Fig. 2b we show off-diagonal element *g* as a function of both the applied static magnetic field *B* and the chemical potential $\mu_c$. The value of g has a large real part in the frequency range of dozens of THz, which becomes more significant when it is applied by a stronger magnetic field. According to equation (4), we can see that the cyclotron frequency has an inverse relationship with the chemical potential, and in equation (2) the most significant magnetic-optic effect occurs around the cyclotron frequency. If we increase the chemical potential, the frequency of the ENZ will be further away from the cyclotron frequency, resulting in a decrease in the value of g.

Fig. 2c and d show the transmission amplitudes of both circularly polarized waves through the composite metamaterial with different amplitudes of applied magnetic field and chemical potential. As the distance increases, we observe that the RCP forward propagating wave almost preserves its amplitude and has only a little attenuation due to the introduction of an imaginary part in the permittivity. In contrast, the LCP propagating wave is almost completely reduced (most of the energy is reflected) and only a small fraction of it tunnels trough the stack. Such a phenomenon of selected circular polarization transmission will become more significant when we increase the applied static magnetic field and decrease the chemical potential.

An incident electromagnetic wave has a linear polarization, which can be decomposed into two opposite circular polarizations, namely, the RCP and LCP propagating waves. Here we assume both circularly polarized waves have unit amplitude and impinge on a graphene/dielectric-stacked layer of thickness 8 $\mu m$. Fig. 3a shows the transmission and reflection spectrum of the RCP component. In the lower frequency, most of the wave is reflected. When it comes to the ENZ region, the structure starts to become almost transparent. When we increase the magnitude of the applied magnetic field, the curves have a red shift due to the stronger magnetic-optic effect. Fig. 3b shows the spectrum of the LCP component, which is similar to the one for the RCP component: at a lower frequency most of the wave is reflected, and the medium becomes transparent when the frequency increases. However, the positions of the turning points are located at higher frequencies as compared with the RCP components. When we increase the magnitude of the applied magnetic field, the spectrum has a blue shift, which has an opposite moving direction to the RCP component. It is obvious that there is an overlapping region between such two groups of spectra, where the RCP component of the wave can transmit through the structure while the LCP component is reflected. Now we define $P_t$ as the product of the transmission intensity of the RCP component and the quality degree of circular polarization, which is defined as $(t_{RCP}^2 - t_{LCP}^2)/(t_{RCP}^2 + t_{LCP}^2)$. In the lower frequency, the transmission light has a high quality of circular polarization but a low intensity, while in the higher frequency, the transmission light has a high intensity but low quality of circular polarization, resulting in a near-zero value of $P_1$ for both sides. Only in the region of ENZ could it have both high transmission intensity and high quality of circular polarization. Fig. 3c shows such a phenomenon. The width of such a peak represents the tolerance of the polarizing effect near the frequency point of ENZ, and has a positive correlation to the magnitude of the applied magnetic field. The relationship is shown in Fig. 3d. The width of the polarizing band has an almost linear relationship with the magnitude of the magnetic field. However, when we decrease the chemical potential while increasing the magnetic field, the polarizing band will become too close to the cyclotron frequency, where the permittivity of the medium shows some dramatic fluctuation, which results in the disorder at the end of the red line in Fig. 3d.

Fig. 4a and b illustrate the situation of oblique incidence. Here we consider point $a_2$ ($b_3$) in Fig. 2a, where the applied magnetic field is 3 T and the chemical potential is 0.35 ev, and the corresponding ENZ frequency is 24.4 THz. The thickness of the composite layer is 8$\mu m$. Fig. 4a and Fig. 4b represent the RCP wave and LCP wave incidence, respectively. According to equation (9), there are four waves mixing in the magnetic-optic material. When we increase the incident angle, it is more likely to excite all four waves in the medium, and thus produce its opposite circular polarization in the transmission spectrum. Both figures show that the graphene/dielectric-stacking polarizer maintains its efficiency at the ENZ region in quite a large incident angle. As the figures show, when the incident angle is less than 45 degrees, the amplitudes of light remain above 90% for both the transmission of RCP incident wave and the reflection of LCP incident wave.

## Conclusion

We theoretically analyzed the transmission and reflection coefficients of both the RCP and LCP waves at the ENZ frequency range of the graphene/dielectric-stacked material, and these results have been further validated by FDTD calculations, which are much more time-consuming. The composite material shows efficient manipulation of the polarization of the electromagnetic wave in a fairly large range of the incidence angle. Both the transmission ratio of the two circularly polarized waves and the working frequency range can be tuned by changing the magnitude of the applied magnetic field or the chemical potential. The suggested structure with the possibility of easy and fast tuning, together with a strong magnetic-optic effect (which is not present in other known materials), may have an impact in a variety of novel devices and applications, beyond circular polarization manipulation and control.

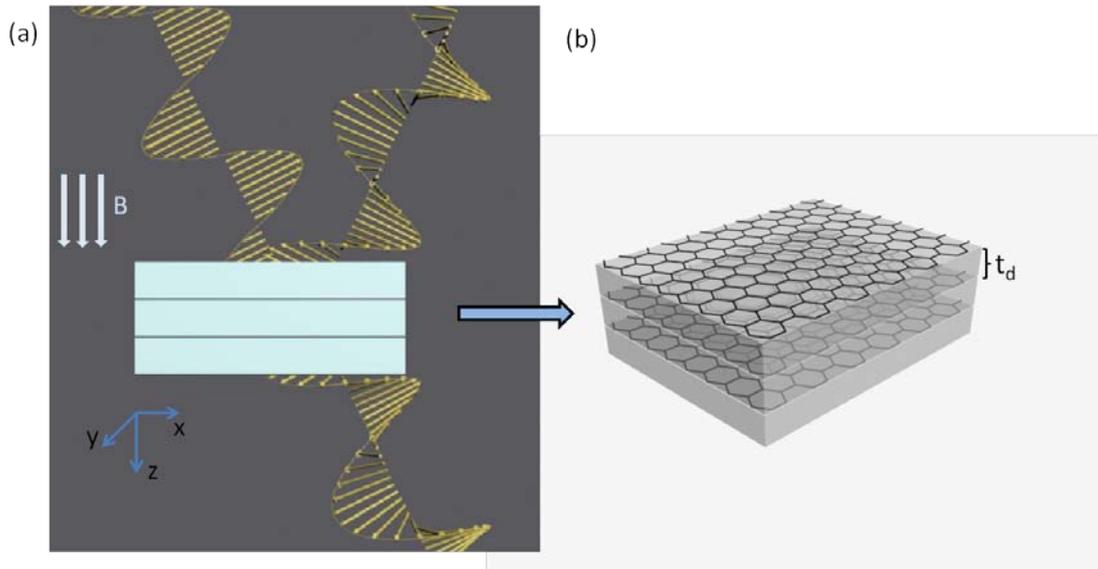

Figure 1(a) Schematic view of the graphene/dielectric-stacked structure for circular polarization control in reflection and transmission. A static magnetic field *B* is applied perpendicularly to the surface. (b) Illustration of our suggested structure.

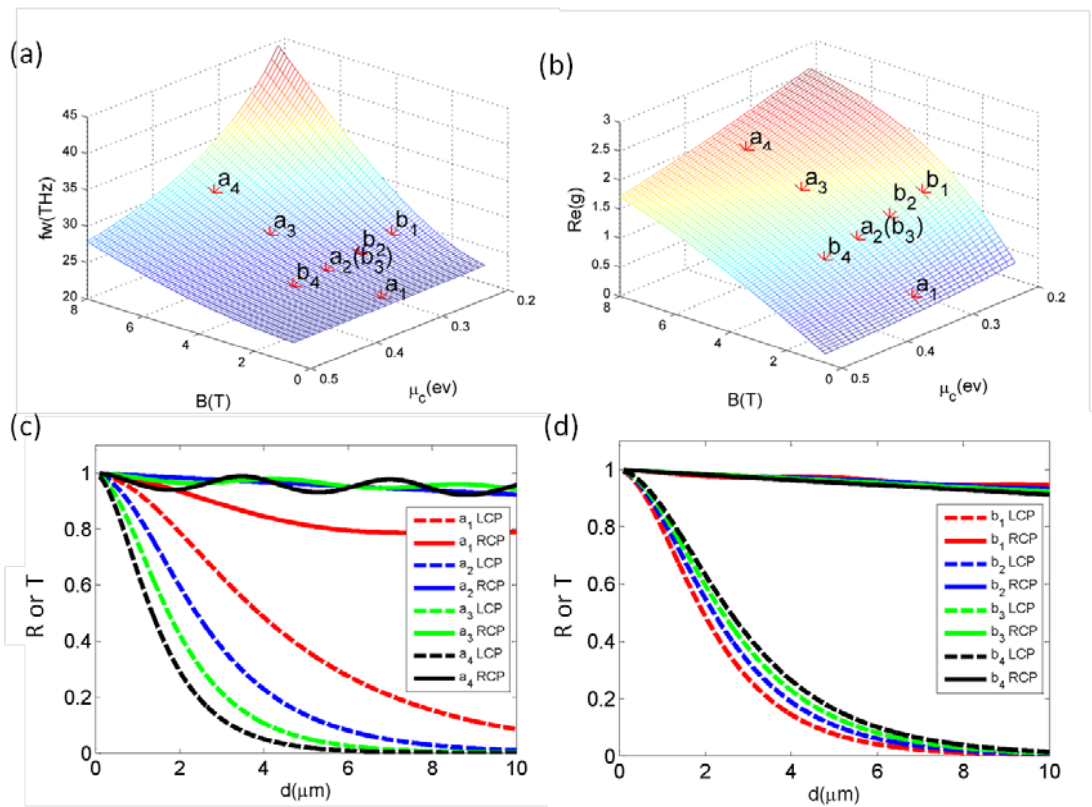

Figure 2.(a) The dependence of the frequency of ENZ with respect to both magnetic field *B* and chemical potential $\mu_c$. (b) The real part of the off-diagonal element $g$ as a function of magnetic field $B$ and the chemical potential $\mu_c$. The thickness of the dielectric layer $t_d$ = *300 nm*, the scattering rate $\Gamma = 2 \times 10^{-3} ev$ and the temperature $T$ = *300 K*. In (a) and (b), the points $a_1$-$a_4$ correspond to the situations where the value of $\mu_c$ is fixed to 0.35 *ev* and increase the magnitude of magnetic field from 1 T to 7 T with a step of 2 T. The points $b_1$-$b_4$ correspond to the situations where the magnitude of B is fixed to 3 T and the chemical potential $\mu_c$ increases from *0.25 ev* to *0.4 ev* with a step of *0.05 ev*. (c) Transmittance of the forward RCP and LCP waves through ENZ graphene/dielectric-stack slab as a function of the slab thickness. The lines correspond to the points $a_i(i=1,2,3,4)$ in (a) and (b) with different applied magnetic fields. (d) Transmittance of the forward RCP and LCP waves as a function of the slab thickness, with a different chemical potential (corresponding to points $b_i$, $i=1,2,3,4$).

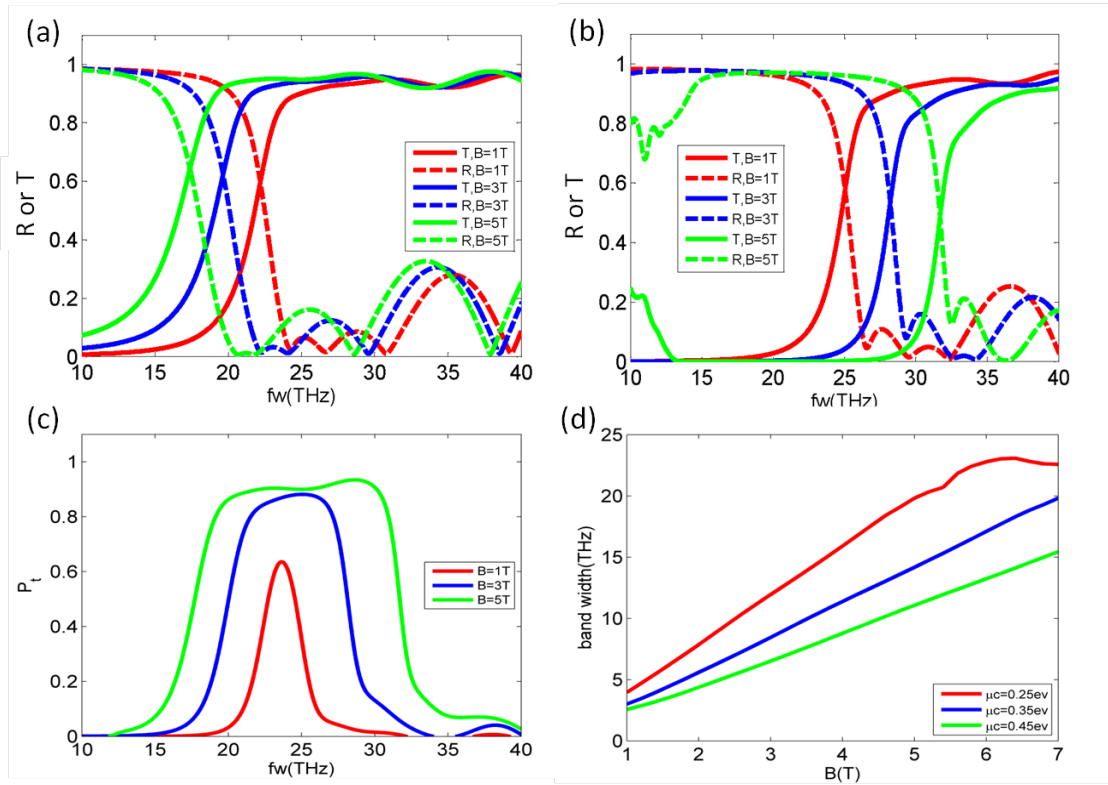

Figure 3. (a) The transmission and reflection spectrum for a forward propagating RCP wave with different applied magnetic fields. (b) The transmission and reflection spectrum for a forward propagating LCP wave with different applied magnetic fields. Both in (a) and (b), the thickness of the graphene/dielectric-stack slab is assumed to be 8 $\mu m$. (c) The efficiency of the polarizer is characterized by the parameter $P_t$, which is defined as the multiplication of the degree of polarization and the strength of the transmission. (d) The bandwidth of the effective operating frequency of the graphene-based polarizer with a dependence on magnetic field *B*.

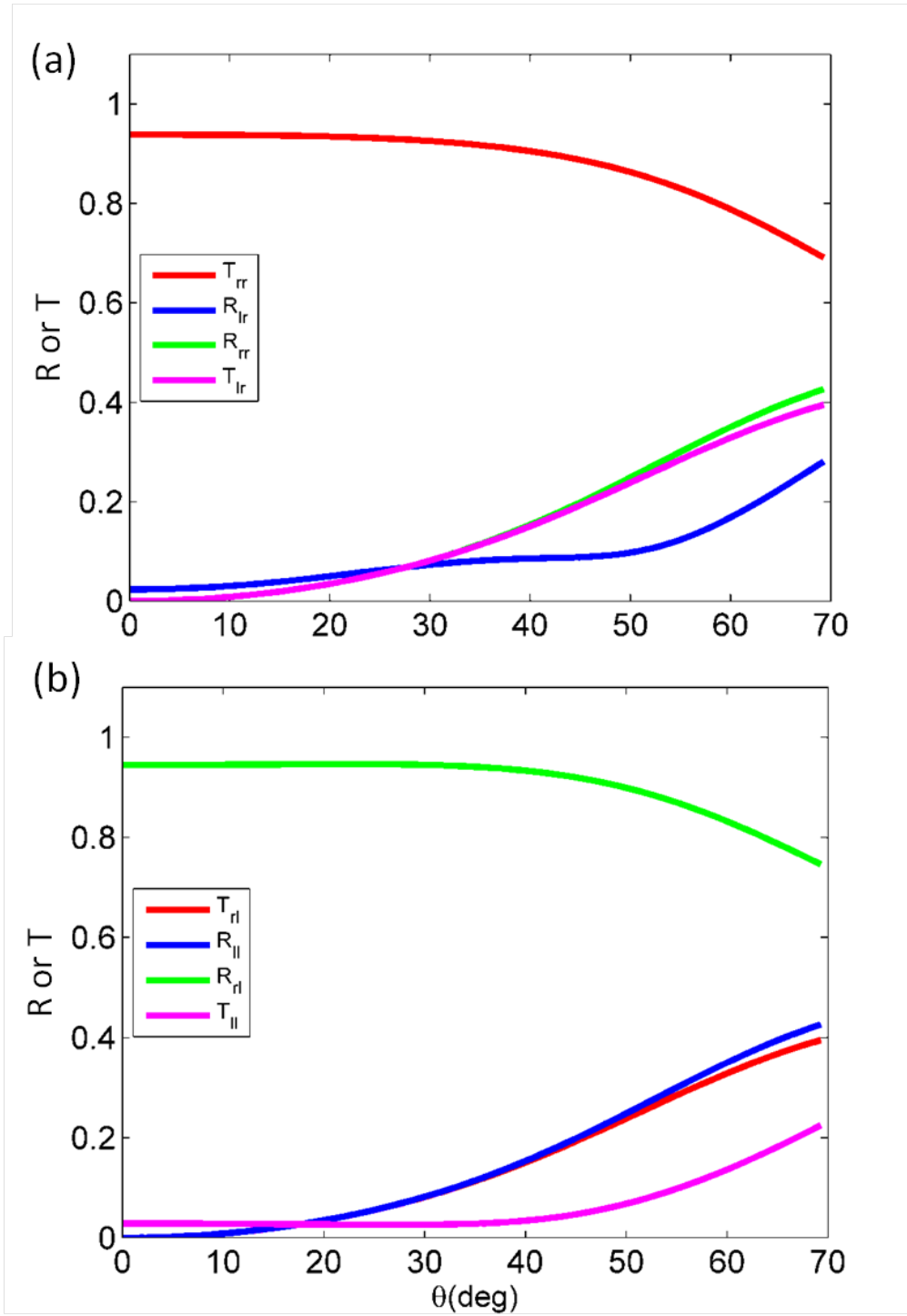

Figure 3. The transmission and reflection of both the RCP and LCP waves with the oblique incidence of the RCP wave (a) and LCP wave (b). $T_{rr}$ is the conversion coefficient from the incident RCP wave to the transmitted RCP wave. $T_{ij}$ and $R_{ij}$ are the conversion coefficients of the circularly polarized waves, e.g., $T_{rr}$ is the conversion coefficient from the incident RCP wave to the transmitted RCP wave, and $R_{lr}$ is the conversion coefficient from incident RCP wave to reflected LCP wave.